# IMPROVING ROUTE DISCOVERY USING STABLE CONNECTED DOMINATING SET IN MANETS

Ramalakshmi R[1] and Radhakrishnan S[2]

[1] Assistant Professor, [2] Senior Professor
[2] Department of Computer Science Engineering, Kalasalingam University, India
`rama,srk@klu.ac.in`

## ABSTRACT

 A Connected Dominating Set (CDS) based virtual backbone plays an important role in wireless ad hoc networks for efficient routing and broadcasting. Each node in the network can select some of its 1-hop neighbors as Multi Point Relay (MPR) to cover all its 2-hop neighbors. A MPR based CDS is a promising approach for broadcasting.  A node in the CDS consumes more energy and the energy depletes quickly than non dominating nodes. Although previous CDS construction algorithms achieve good results in terms of the size of CDS, a minimum size CDS does not necessarily guarantee an  optimal network performance from an energy efficient point of view. In this paper, we propose a distributed algorithm for energy efficient stable MPR based CDS construction to extend the lifetime of ad hoc wireless networks by considering energy and velocity of nodes. We have also implemented route discovery protocol to make use of the CDS nodes to relay route request messages. The simulation results show that our algorithm increases the lifetime up to 25% than previous works and 60% reduction in the route request messages during route discovery process.

## 1. INTRODUCTION

Wireless ad hoc networks are self configuring networks, can be deployed for many applications such as automated battlefield, search and rescue and disaster relief.  Mobile ad hoc network (MANET) consists of wireless nodes that communicate with each other without any infrastructure. A communication session is achieved either through a single hop radio transmission if the communication parties are within the transmission range, or through relaying by intermediate nodes otherwise. Two important features of an ad hoc network are its dynamic topology and resource limitation.  These features make routing decision very challenging. Every node in mobile ad hoc networks can move in any direction at any time and any speed. A temporary infrastructure or a virtual backbone can be formed to provide communication. This virtual backbone may be broken due to the node movement.

A network can be modeled as a Unit Disk Graphs (UDG) where two nodes are connected if they are within each other's transmission range. To support various network functions, some wireless nodes are selected to form a virtual backbone. It is proved that multipoint relaying (MPR) is an efficient stated for on-the-fly broadcasting in Mobile Ad Hoc Networks. The relaying nodes which are selected by the source node are responsible for flooding a receiving packet. A connected dominating set (CDS) consists of all the relay nodes.  Finding connected dominating set is a NP-hard problem. There are many algorithms based on MPR to reduce the size of the CDS. In the original MPR based CDS schemes [1, 2, 11, 12, 13], nodes are chosen based on node id or node degree. The CDS selection has to consider other information such as energy, bandwidth and mobility in order to provide suitable links for some specific applications.

In this paper, we propose an approach to generate quality connected dominating set based on MPR by considering the energy and link velocity factors. There are many works on MPR based CDS construction [1, 2, 11, 12, 13, 14], but none of work considered the energy of nodes and their mobility together for the construction work.

## 1.1 Problem Definitions

The connected dominating set or virtual backbone is proposed to facilitate routing, broadcasting and establishing a dynamic infrastructure. Minimizing the CDS produces a simpler abstracted topology of the MANET and allows for using shorter routes between any pair of hosts. A wireless network is usually modeled as a unit disk graph G = (V, E) where V is the set of nodes and E is the set of links in the network. Each node has uniform transmission range R. Each node in V is associated with coordination in 2-D Euclidean space. A wireless link (u,v) ∈ E if and only if the Euclidean distance between nodes $u$ and $v$ is smaller than transmission range R.

**Connected Dominating set:**
A subset S ⊆ V is called a dominating set (DS) of G if {$\forall v \in V$, either $v$ is in S or has an adjacent neighbor in S}. The subset S is called Connected Dominating Set if the graph G', induced by S is connected i.e, G'[S] is connected.

**Multipoint Relay:**
For a given a graph G=(V,E) and a node $v \in$ V, let $N_1(v)$ and $N_2(v)$ represent the set of 1-hop and 2-hop neighbors of v, respectively. MPR asks for a minimum size subset MPR of $N_1(v)$ such that $N_2(v)$ is covered by MPR.

The rest of the paper is organized as follows: Section 2 describes the works related to CDS construction; Section 3 describes the energy efficient stable connected dominating set construction algorithm and route discovery process using CDS nodes. We describe our simulation environment and the performance metrics in Section 4. Section 5 concludes the paper.

## 2. RELATED WORKS

There are many work on MPR based CDS construction for efficient broadcasting in Manet. Adjih et al [12] proposed a source independent MPR called the MPR-CDS. The algorithm starts by having every node $v$ calculate its source dependent MPR. After that every node $v$ decides whether it belongs to the MPR-CDS or not according to the following simple rules:

- Rule 1: Node u ∈ MPR-CDS iff v has the smallest ID in its 1-hop neighborhood.
- Rule 2: Node v ∈ MPR-CDS iff v ∈ w's MPR where w's ID is the smallest in v's 1-hop neighborhood.

In [1] Wu, has noticed that in many occasions nodes added by rule-1 of MPR-CDS algorithms are useless. Moreover the algorithm used to calculate the source dependent MPR does not benefit from Rule-2 of the MPR-CDS algorithm. In [1] Wu modified Rule 1 as follows:

- **EMPR:** node v ∈ MPR-CDS iff *v* has the smallest ID in its 1-hop neighborhood and v has at least two unconnected neighbors.

Moreover, Wu modified the MPR calculation algorithm in [12] by having every node v start by adding all its free neighbors to its MPR set. A node u is a free neighbor of node v iff u ∈ N(v) and v is not the smallest ID neighbor of u.

Chen et al [11] observed that the node degree is more related to the size of a CDS than the node ID and three improvements are proposed. They replaced the EMPR rule with two rules based on degree called DEMPR.

- Rule 1: node v ∈ MPR-CDS if v has the largest node degree among all its one-hop neighbors and v has two unconnected neighbors.
- Rule 2: node v ∈ MPR-CDS if v has been selected as an MPR and its selector has the largest node degree among its one-hop neighbors.

Badis et al [13], proposed heuristic referred to as the QoS based MPR-1(QMPR-1) follows the same steps as the original MPR heuristic but it modifies the tie breaking procedure. Instead of a maximum node degree, a node with high bandwidth is chosen when multiple choices exist.

There are few works on energy efficient CDS construction. In [6] Kim extended the Mac-layer timer based connected dominating set protocol by considering energy level at each node to construct energy aware CDS. In [7], Ruiyun Yu proposed an energy efficient dominating tree construction (EEDTC) algorithm with two phases, marking phases followed by connection phase. In the marking phase, a Maximal Independent Set (MIS) is constructed and connectors are added to make it as CDS. In [14], Wu proposed a method to calculate power aware connected dominating set. They used degree and residual energy level of nodes to reduce the CDS size to prolong the lifespan of the nodes in the CDS.

Only few works are done in stable connected dominating set construction. In [15], Change proposed Dynamic Power-aware and Stability-aware Multipoint relays which avoid selecting the border nodes as the forwarding nodes. They used power adaptive broadcasting by reducing the transmission range of mobile nodes to save energy. The range buffer based approach is proposed to further enhance the stability of the forwarding nodes. In [4], Meganathan proposed an algorithm to determine stable connected dominating set based node velocities. Their algorithm prefers slow moving nodes with lower velocity rather than the usual approach of preferring nodes with a larger number of uncovered neighbors. They compared their method with another work which is based on node degree.

In [17], dominant pruning rules are applied to distribute the route request messages in AODV protocol. Authors have proposed an algorithm for Two Hop Horizon Pruning in [16] to distribute route request messages using 2-hop dominating set. They have applied virtual radio range and radio range to remove some of the neighbors from 1-hop neighbor list.

Although previous CDS construction algorithms achieve good results in terms of the size of the CDS, a minimum size CDS does not necessarily guarantee the optimal network performance from an energy efficient point of view. This motivated us to construct an energy efficient stable connected dominating set construction to prolong the network lifetime.

## 3. ENERGY AWARE STABLE CONNECTED DOMINATING SET CONSTRUCTION ALGORITHM (EAS-CDS)

### 2.1. Notations and Assumptions

We assume that every node in the network has same transmission range R. Two nodes are connected if the Euclidean distance between the nodes is less than R. We used the notations for our algorithm as in Table 1.

**Table 1 Notations**

| | |
|---|---|
| $N_1(u)$ | Open neighbor set of node u |
| $N_1[u]$ | Closed neighbor set of node u N[u] = N(u) $\bigcup$ {u} |
| $N_2(u)$ | 2-hop neighbor set of node u |
| $Erg_x$ | remaining residual energy at node x |
| $Vel_x$ | Velocity of node x |
| MPR(u) | MPR set of node u |
| $Deg_u$ | No of neighbors of u |

### 2.2. EES-CDS Algorithm

Our algorithm for energy aware stable connected dominating set construction (EAS-CDS) consists of three phases: Neighbor Discovery Phase, CDS Formation Phase and pruning phase. During the Neighbor Discovery Phase, there is an initial exchange of messages via which a node $u$, made aware of its $N_2(u)$. In the CDS Formation Phase, a node $u$ locally selects a set MPR($u$) of its $N_1(u)$ as its multipoint relays by using simple greedy algorithm and in the pruning phase, two rules are applied to reduce the connected dominating set size.

In [1], Wu proposed a simple decentralized algorithm for the formation of connected dominating set in ad hoc networks. This algorithm is based on marking process. We have used their algorithm to choose the multipoint relays for a node. In [14], Dai proposed two rules for power aware CDS construction using node degree and residual energy level of nodes. In this work, we have modified the rules proposed by [14] with energy level and velocity to prolong the stability and to reduce the size of a connected dominating set generated from the marking process.

Let the graph induced by CDS be G'.

**Rule 1**: Consider two marked vertices $v$ and $u$ in G'. The marker of $v$ is changed to gray if one of the following conditions holds:

i)   $N[v] \subseteq N[u]$ in G and $Erg_v < Erg_u$

ii)  $N[v] \subseteq N[u]$ in G and $Vel_u > Vel_u$ when $Erg_v = Erg_u$

iii) $N[v] \subseteq N[u]$ in G and $Deg_v < Deg_u$ when $Erg_v = Erg_u$ and $Vel_v = Vel_u$

The above rules indicate that when the closed neighbor set of $v$ is covered by that of $u$, vertex $v$ can be removed from G' if the energy level of $v$ is smaller than u. Velocity is used to break a tie when energy levels of $u$ and $v$ are same. Degree is used to break the tie if both energy levels and velocity of $u$ and $v$ are same. Node ID can be used to break the tie, in case all the values are same.

**Rule 2**: Assume that $u$ and $w$ are two marked neighbors of marked vertex $v$ in G'. The marker of $v$ can be changed to Gray if one of the following conditions holds:

1. $N(v) \subseteq N(u) \bigcup N(w)$, but $N(u) \not\subset N(v) \bigcup N(w)$ and $N(w) \not\subset N(u) \bigcup N(v)$ in G

2. $N(v) \subseteq N(u) \bigcup N(w)$ and $N(u) \subseteq N(v) \bigcup N(w)$, but $N(w) \not\subset N(u) \bigcup N(v)$ in G and one of the following conditions holds:

    a. $Erg_v < Erg_u$ or

    b. $Erg_v = Erg_u$ and $Vel_v > Vel_u$ or

    c. $Erg_v = Erg_u$ and $Vel_v = Vel_u$ and $Deg_v < Deg_u$

3. $N(v) \subseteq N(u) \bigcup N(w)$ and $N(u) \subseteq N(v) \bigcup N(w)$, but $N(w) \subseteq N(u) \bigcup N(v)$ in G and one of the following conditions holds:

    a. $Erg_v < Erg_u$ and $Erg_v < Erg_w$ or

    b. $Erg_v = Erg_u < Erg_w$ and $Vel_v > Vel_u$ or $Deg_v < Deg_u$ when $Vel_v = Vel_u$

    c. $Erg_v = Erg_u = Erg_w$ then

        i. $Vel_v > Vel_u$ and $Vel_v > Vel_w$ or

        ii. $Vel_v = Vel_u > Vel_w$ and $Deg_v < deg_u$ or

        iii. $Vel_v = Vel_u = Vel_w$ and $Deg_v = min\{Deg_v, Deg_u, Deg_v\}$

The above rule indicates that when v is covered by u and w; the conditions are

1) if neither $u$ or $w$ is covered by the other two among $u$, $v$ and $w$ then $v$ is unmarked

2) if neighbor set of $v$ is covered by $u$, $w$ and neighbor set of $u$ is covered by $v$, $w$ but neighbors of $w$ are not covered by $u$, $v$ then $v$ is unmarked if the energy level of $v$ is smaller than $u$. Velocity is used to break the tie if energy levels are same. Degree is used to break the tie if both energy and velocity values are same.

3) If neighbor set of $v$, $u$ and $w$ are covered by the other two among $u$, $v$ and $w$ then node $v$ is unmarked with conditions: energy level of $v$ is less than $v$ and $w$; the energy level of $v$ is same as $u$ but smaller than $w$, velocity value is used for unmarking. If velocity values are same then degree of $v$ is used to break the tie.

The procedure for the energy efficient CDS construction algorithm is given in Table2.

It is clear that node $u$ only has to wait for the information about its two hop neighbors. The set of all MPR is a connected dominating set of the entire network. The node terminates the construction phase by communicating its final decision to all its neighbors. The result of this algorithm for a sample network with 10 nodes is given in figure 1. The MPR of node 1 is 2 to cover its 2-hop neighbor 10. The 2-hop neighbors of node 2 are {4, 5, 8, and 9} and its MPR is 10. Likewise, MPR(9)=7 and MPR(10)=9. Thus the final CDS consists of nodes {2, 10, 9, and 7}.

## Table 2: Algorithm for Energy Aware Stable CDS Construction

**Algorithm: EAS-CDS**

   **Input**:  An undirected graph G(V, E)

   **Output**:  Minimum Size CDS which is energy efficient and Stable.

- **Neighbor Discovery phase**

   Nodes periodically exchange *hello* messages for neighbor discovery.  Every node sends and receives *hello* messages but does not forward them.  A *hello* message generated by a node u contains its ID, Energy ($Erg_u$), Velocity ($Vel_u$) and list of neighbors N(u). This one hop neighborhood exchange enables every node *u* to obtain its two hop neighborhood information.

- **CDS Formation Phase**

   MPR Selection

      i. Initially all nodes are in White colour

      ii. Every node *v* assigns its $N_1(v)$ to MPR(*v*) if it has two unconnected neighbors.

      iii.   Mark all the nodes in MPR(*v*) to Black Colour

      iv.   Marks all the neighbors of MPR(*v*) to Gray Colour

- **Pruning Phase**

   Apply Proposed Rule 1 and Rule 2 to all dominating nodes.

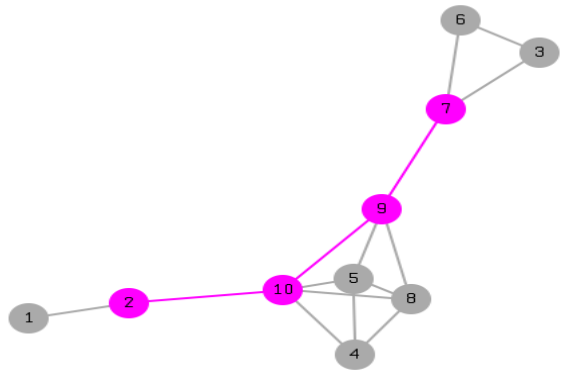

Figure 1: CDS = {2, 7, 9, and 10}

## 2.3. Route Discovery using EES-CDS

On demand routing protocols like AODV, DSR for ad hoc networks use route discovery process to find the path between source and destination. In [18], the source node initiates the route discovery when it has no route to the destination. It broadcasts a route request packet (RREQ) to its neighbors. Each receiving node in turn broadcasts RREQ packet. This process is repeated until the packet reaches the destination and the destination node will send the route reply message (RREP) to the source. This type of route discovery leads to broadcast storm problem. To overcome this problem, we implemented the route discovery process in AODV using the CDS nodes only. When a CDS node receives a RREQ packet, it broadcasts the packet and non-CDS nodes only receive the packet. The non-CDS are not rebroadcasting the RREQ packets. Thus the number of RREQ packet transmission is reduced and the network congestion is avoided.

## 4. SIMULATION RESULTS AND ANALYSIS

### 4.1. Simulation Environment

We implemented our algorithm EAS-CDS in ns-2.34. To evaluate the performance of our algorithm, we also implemented the approach proposed by Wu in [1] and Meganathan in [4]. In this section, the simulation results are reported and analyzed. To generate a network, $n$ nodes are randomly placed in a 1000m x 1000m region. Each node has uniform transmission range 250m and is associated with an initial energy values from 1J to 15J. In our simulations, the number of nodes $n$ has been assigned the values 50, 100, 150, 200, 250. This allows us to test our algorithm from sparse to dense networks. Any two nodes distance less than the transmission range are considered neighbors. Each node moves randomly in this area with a speed in the range [0 .. $V_{max}$] and pause time of 100s. The values of $V_{max}$ are 5, 10 and 25m/s. Each simulation is conducted for 600s and it is repeated 10 times. The parameters used in our simulation are listed in Table 3.

| Table 3 : Simulation Parameters | |
|---|---|
| Network Area | 1000m$^2$ |
| Number of Nodes | 50..250 |
| Transmission Range | 250m |
| Mobility Speed | 5m/s, 15m/s, 25m/s |
| Initial Energy | 1J..15J. |
| Energy for transmission | 1.4W |
| Energy for Receiving | 1.0W |
| Idle energy | 0.013W |
| Pause time | 100s |
| Simulation Time | 600s |
| Propagation Model | Two-ray Ground |
| MAC | IEEE 802.11 |
| Antenna | Omni Antenna |
| Mobility Model | Random Way Point |
| Data Traffic | CBR packets with 512 Bytes |
| No of Sources | 20 |
| Packet Sending Rate | 5 packets /sec |

We implemented this algorithm by using three messages: The first message is for exchanging the list of neighbors, the second is used by a node *u* to communicate to a neighbor *v* for which *u* is the MPR selector and the final one is used by a node to make every neighbor aware of its final decision.

### 4.2. Result Analysis

We measured the performance of our work in terms of

- CDS Size

It describes the no of nodes included in the CDS to act as broadcast relay nodes. Figure 4.1 shows the average no of nodes included in the CDS with different mobility values 5m/s, 15m./s and 25m/s. The results show that CDS generated by our algorithm is larger than Wu [1] and less than Meganathan [4]. In [1], Wu used node degree for selection and Meganathan in [4] selected nodes with only lower velocity. The average size of the CDS increases with network density.

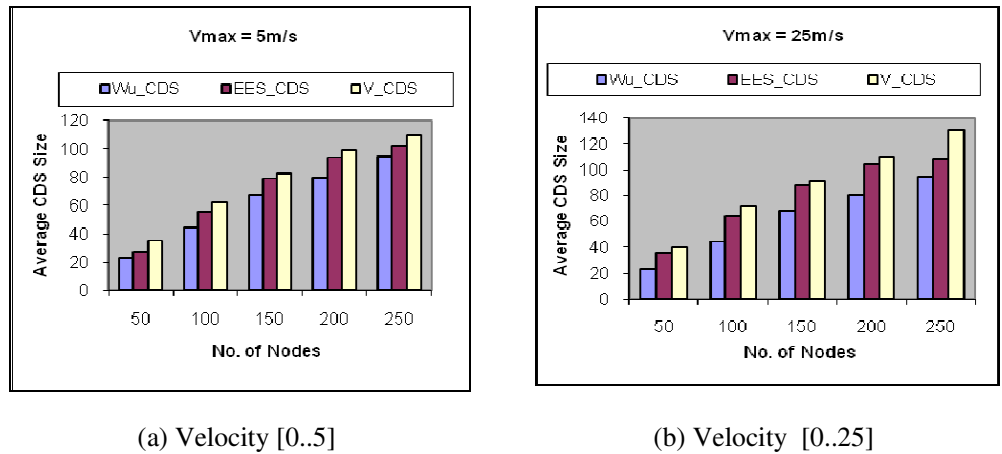

(a) Velocity [0..5]                    (b) Velocity  [0..25]

**Figure 4.1 CDS Size**

- CDS Stability

The simulation stops when the energy level of at least one node becomes 0. Figure 4.2 shows comparison of our work with Wu [1] and Meganathan [4] in terms of lifetime. Our work outperforms well than the other two works because nodes in the CDS generated by our algorithm has high energy level and minimum velocity. The CDS stability is high when the node velocity is low. In [4], priority is given to slow moving nodes but they don't consider the energy level of the nodes.

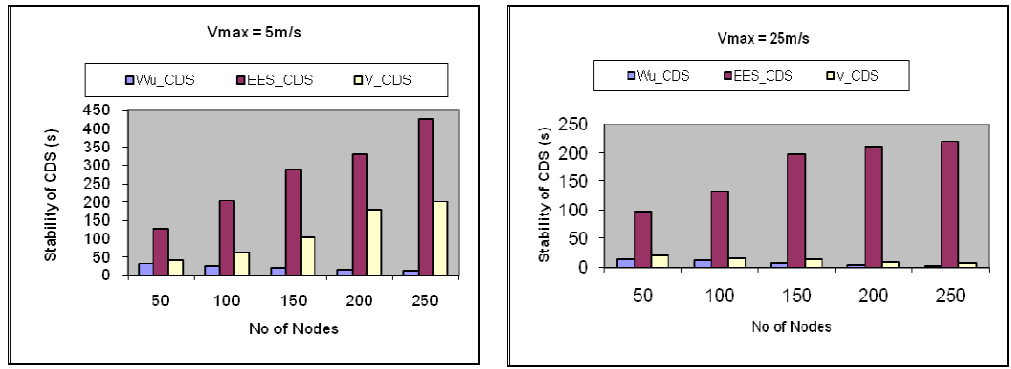

(a) Velocity [0..5 ]                                    (b) Velocity [0..25 ]

**Figure 4.2 CDS Lifetime**

- Packet Delivery Ratio

We have implemented the route discovery process of AODV using only CDS nodes. As in
Figure 1, only nodes {2, 7, 9, and 10} will relay the RREQ packets and other nodes will receive
the packet, they don't broadcast. We tested this route discovery process with different network
size. We generated 100 CBR packets of 512 bytes per second with pause time 100s. The AODV
protocol uses simple flooding mechanism, so the no of packet transmission is high. But, only
CDS nodes are used for RREQ transmission, the number of RREQ transmission is minimum than
AODV. The results are given in Figure 4.3. The packet delivery ratio is given in Figure 4.4. We
assumed 30 sources with different packet flows. The results show that the delivery ratio is high
when packet flow is less, otherwise the performance is close to AODV.

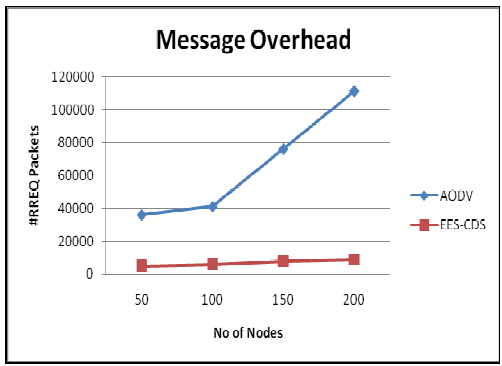          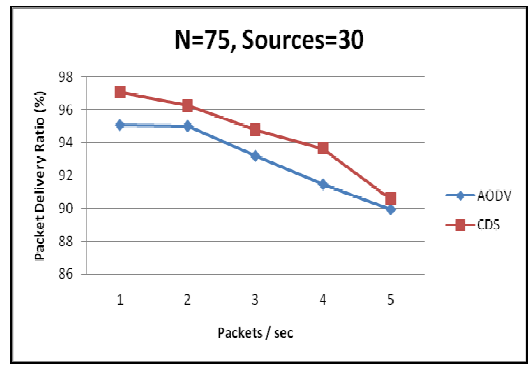

**Figure 4.3 #RREQ Transmission**          **Figure 4.4 Packet Delivery Ratio**

## 5. CONCLUSION AND FUTURE WORK

In this work, we have proposed an algorithm for power efficient stable connected dominating set
construction. We have proposed two rules to reduce the connected dominating size and to prolong
the life span of the nodes with residual energy level and velocity. We have also implemented the
route discovery process of AODV using CDS nodes. The results from the simulation show that
our work outperforms in terms of stability and it increases the lifetime by 25% to 30%. The route discovery process over CDS reduces the route request transmission by 60%. We plan to verify the performance of OLSR protocols using our MPR nodes as relays.

## ACKNOWLEDGEMENT

The authors would like to thank the Project Coordinator and Project Directors of TIFAC-CORE in Network Engineering, Kalasalingam University for providing the infrastructure facilities in Open Source Technology Laboratory and also thank Kalasalingam Anand Ammal Charities for providing professional development allowance for this work.

## References

[1]   J.Wu & W. Lou, (2006) "Extended Multipoint Relays to Determine Connected Dominating Sets in MANETs", IEEE *Transaction on Computers*, vol. 55, pp334-347.

[2]   J.Wu, (2003) "An Enhanced Approach to Determine a Small Forward Node Set Based on Multipoint Relay", *Proc. IEEE Semi-Ann. Vehicular Tech. Conference.*

[3]   N.Meganathan, & A.Farago, (2008) "On the Stability of Paths, Steiner Trees and Connected Dominating Sets in Mobile Ad Hoc Networks", *Ad hoc Networks*, Vol. 6, pp744-769.

[4]   N.Maganathan, (2010) "Use of Minimum Node Velocity Based Stable Connected Dominating Sets for Mobile Ad hoc Networks", *IJCA special issues on "Mobile" Ad-Hoc Networks*, 89-96.

[5]   F.Dai & J.Wu, (2004) "An Extended Localized algorithms for Connected Dominating Set Formation in Ad Hoc Wireless Networks", *IEEE Trans. Parallel and Distributed Systems,* Vol. 15.

[6]   B. Kim,D.Zhou, J.Yang & M.Sunl, (2005)  "Energy-Aware Connected Dominating Set Construction in Mobile Ad Hoc Networks," *Auburn University Technical Report*, CSSE05-07, 2005

[7]   Ruiyun Yu, Xingwei Wang, & Sajal K.Das, (2009) "EEDTC: Energy-Efficient Dominating tree Construction in Multi-hop Wireless Networks", *Pervasive and Mobile Computing,* Vol. 5, pp318-33.

[8]   S. Funke, A. Kesselman, U. Meyer, & M. Segal. (2006) "A simple improved distributed algorithm for minimum   Connected dominating set in unit disk graphs*",  ACM Transactions on Sensor Networks,* Vol. 2, pp444–453.

[9]   P.Wan, K K.M. Alzoubi & O. Frieder, (2002) "Distributed construction of connected dominating set in wireless ad hoc networks*", Proc. 21th Annual Joint Conference of the IEEE InfoCom.*

[10]  Y.Wu, F. Wang, M. T. Thai, & Y. Li, (2007)  "Constructing k-connected m-dominating sets in wireless sensor networks", *Proc. Military Communications Conference.*

[11]  X.Chen & J.Shen, (2004) "Reducing Connected Dominating Set Size with Multipoint Relays in Ad Hoc Wireless Networks", *Proc 7th Int. Sym. Parallel Architectures, Algorithms and Networks*, pp539-43.

[12]  C.Adjih, P.Jacquet, & L.Viennot, (2002) "Computing Connected Dominating sets with multipoint relays", *Technical Report*, INRIA.

[13]  Hakim Badis, (2004)  "Optimal Path Selection in a Link State QoS Routing Protocol", *Proc. IEEE VTC2004* Spring.

[14]  Wu J, Dai F, Gao M, & Stojmenovic I, (2002) "On calculating Power Aware Connected Dominating Sets for Efficient Routing in Ad Hoc Wireless Networks", *Journal of Communication and Networks*, vol.4, No.1, pp1-12

[15] Change Y, Ting Y, & Wu S, (2007) "Power-Efficient and Path-Stable Broadcasting scheme for Wireless Ad Hoc Networks", *Int. Conf. Advanced Information Networking and applications Workshops (AINAW'07).*

[16] Marco Aure´lio Spohn & J.J. Garcia-Luna-Aceves (2006), "Improving route discovery in on-demand routing protocols using two-hop connected dominating sets", *Ad Hoc Networks*, Vol. 4, pp509-531.

[17] Marc Mosko, J.J. Garcia-Luna-Aceves & Charles E. Perkins (2003), "Distribution of Route Requests Using Dominating-Set Neighbor Elimination in an On-demand Routing Protocol", GLOBECOM 2003.

[18] Belding-Royer E, Perkins C & Das S, (2003), "Ad hoc on-demand distance vector routing", *Internet Draft www.ietf.org/rfc/rfc3561.txt*

## Authors

**RAMALAKSHMI R** received her Master of Engineering degree from Anna University, Chennai. She is doing her doctoral program. She is a member of CSI, ISTE and Network Technology group of TIFAC-CORE in Network Engineering.  Her areas of interest include Wireless Sensor Networks, Mobile Ad Hoc Networks and Graph Theory applications in ad hoc networks.

**RADHAKRISHNAN S** received his Master of Technology and Doctorate in Philosophy from Banaras Hindu University, Banaras. He is the Senior Professor in Computer Science Department of Kalasalingam University, Srivilliputhur. He is a member of ISTE. He is the Project Director of Network Technology Group of TIFAC-CORE in Network Engineering.  He has produced 12 Ph.D's and currently guiding research scholars in the areas of Network Security, Sensor Networks, Evolutionary optimization, Mesh Networks and Cloud Computing.